\documentclass[twocolumn,nofootinbib,preprintnumbers,amsmath,amssymb]{revtex4}

\usepackage{graphicx}% Include figure files
\usepackage{dcolumn}% Align table columns on decimal point
\usepackage{bm}% bold math
\usepackage{amsmath}

\def \be {\begin{equation}}
\def \ee {\end{equation}}

\begin{document}

\title{Distances and moments of inertia of Fermi Pulsars}% Force line breaks with \\

\author{Andrei Gruzinov}

\affiliation{ CCPP, Physics Department, New York University, 4 Washington Place, New York, NY 10003
}

\begin{abstract}

Measurement of distances and moments of inertia of pulsars must be useful, for instance, for three-dimensional mapping of the dispersion and rotation measures, constraining the nuclear equation of state, etc. The distances and moments of inertia can be measured by fitting the gamma-ray lightcurves of pulsars, because the gamma-ray emission seems to be governed by easily calculable physics. The (first-principle) theoretical lightcurves have been computed only for weak pulsars (pair production near the light cylinder much smaller than Goldreich-Julian \cite{Goldreich1969} per rotation), and at insufficient accuracy; but, since this computation has been done by a self-taught numericist, it must be possible to improve the theoretical accuracy for weak pulsars, and also to extend the computation to non-weak pulsars.

To invite the computational effort of better-equipped researchers, we describe  an (entirely obvious) procedure for measuring the distances and moments of inertia for weak pulsars.

\end{abstract}

\maketitle

\section{Distances and moments of inertia}\label{distances}

Let us describe (an highly idealized) measurement which we expect to work for weak pulsars. A proper introduction, \S\ref{intro}, and discussion of the theory, \S\ref{arist}, follow. 

From the Fermi catalog \cite{Fermi2013} we take the following:
\begin{itemize}

\item period $P$;
\item period derivative $\dot{P}$;
\item bolometric flux $f$ ($[f]={{\rm erg}\over {\rm cm}^2{\rm s}})$;
\item photon energy cutoff $E_{\rm cut~obs}$;
\item bolometric lightcurve $l_{\rm obs}(\phi )$, where $\phi$ is the pulse phase; $l_{\rm obs}(\phi )$ is proportional to the bolometric flux at a given phase, and normalized by $l_{\rm obs~max}=1$.

\end{itemize}

For weak pulsars, the 'theory' \cite{Gruzinov2013} gives the following:
\begin{itemize}

\item Photon cutoff energy is 
\be\label{cut}
E_{\rm cut~th}=e(\theta , \chi)L_{34}^{3/8}P_{\rm ms}^{-1/4}{\rm GeV}.
\ee
Here $\theta$ is the spin-dipole angle, $\chi$ is the observer angle (the angle between the spin axis and the direction to observer), $L_{34}$ is the spin-down power in units of $10^{34}$erg/s, $P_{\rm ms}$ is the period in ms. The dimensionless function $e(\theta , \chi)$ is currently known to some 10\% accuracy for an axisymmetric pulsar ($e(0, \chi)$ drops from about $5$ at $\chi \approx 90^\circ $ to about $2.5$ at $\chi \approx 65^\circ $) and to yet unclear accuracy for generic $\theta$.

\item The normalized bolometric lightcurve is 
\be\label{curve} 
l_{\rm th}(\phi )=l_{\rm th}(\phi ; \theta ,\chi).
\ee
The accuracy of $l_{\rm th}(\phi ; \theta ,\chi)$ is yet unclear.

\item Bolometric efficiency
\be\label{eff} 
\epsilon =\epsilon(\theta ,\chi),
\ee
defined as the ratio of the pulsed bolometric luminosity (as seen at observation angle $\chi$) to the spin-down power. The dimensionless function $\epsilon(\theta ,\chi)$ is currently known to some 10\% relative accuracy for an axisymmetric pulsar ($\epsilon (0, \chi)$ drops from about $10$ at $\chi \approx 90^\circ $ to about $1$ at $\chi \approx 85^\circ $, to about $0.1$ at $\chi \approx 65^\circ $) and to yet unclear accuracy for generic $\theta$.

\end{itemize}

The measurement procedure is then straightforward: 

\begin{itemize}

\item Use Eq.(\ref{curve}) to fit the lightcurve, thereby measuring both $\theta$ and $\chi$.

The lightcurves (both observational and theoretical) are rich enough (at least in some cases with many local maxima, etc.), and this may work.

\item Use Eq.(\ref{cut}) to measure the spin-down power $L_{34}$, and then, knowing $P$ and $\dot{P}$, deduce the moment of inertia of the pulsar.

\item Use Eq.(\ref{eff}) to calculate the bolometric luminosity, and then, knowing the bolometric flux $f$, deduce the distance to the pulsar.

\end{itemize}

\section{Pulsar Theory}\label{intro}

A first-principle, i.e., using no arbitrary parameters, computation of pulsar spectra and lightcurves has been presented in \cite{Gruzinov2013}. The theoretical results are supposedly exact, or close to exact, at least in principle (although the current numerical accuracy is poor). 

Only weak pulsars have been treated. The Fermi pulsar catalog \cite{Fermi2013} supposedly contains many weak pulsars; and perhaps most Fermi pulsars can be usefully approximated as weak (for strong pulsars, the averaged efficiency must be much smaller than the Fermi's median value of about 15\%). 

The non-weak pulsar problem (significant, as compared to Goldreich-Julian per rotation, pair production near the light cylinder) does not seem to be insurmountably more difficult either, and is expected to be solved in the near future.

Once the pulsar theory delivers, the procedure outlined in \S\ref{distances} will become feasible. A logical questions is why don't we do it here. We are currently computing a library of magnetospheres with different spin-dipole angles $\theta$, and we will attempt the lightcurve fitting. The results will be published, regardless of whether we fail or succeed. (It is already clear from \cite{Gruzinov2013} that we cannot fail too miserably.) 

We want to stress that the core of the theory, Aristotelian Electrodynamics (AE, \S\ref{arist}), appears, by virtue of near-triviality, to be  unassailable; our potential failure can only come from bad numerics and/or failure of the calculation recipe \cite{Gruzinov2013} and/or non-weak pulsar effects. These problems, if they indeed occur, should be temporary. It seems very likely that AE is capable of fully solving the pulsar (only at high energies, of course). The purpose of this note is to invite computational effort of other researchers.

\section{Aristotelian Electrodynamics}\label{arist}
AE (numerical) calculation of the pulsar gives the electromagnetic field and positron and electron densities everywhere in the magnetosphere. The electromagnetic field is computed, starting from zero, by Maxwell equations. To solve Maxwell equations, one needs to know the electric current. 

The electric current inside the star is given by the standard Ohm's law (plus permanent current responsible for the magnetization). The electric current outside the star is 
\be\label{cur}
{\bf j}=\rho_+{\bf v}_+-\rho_-{\bf v}_-,
\ee
where $\rho_\pm$ and ${\bf v}_\pm$ are the (positron charge normalized) number densities and velocities of positrons and electrons. 

In AE, 
\be\label{ae}
{\bf v}_{\pm}={{\bf E}\times {\bf B}\pm(B_0{\bf B}+E_0{\bf E})\over B^2+E_0^2}.
\ee
Here the scalar $E_0$ and the pseudoscalar $B_0$ are the proper electric and magnetic fields defined by 
\be
B_0^2-E_0^2=B^2-E^2,~ B_0E_0={\bf B}\cdot {\bf E},~ E_0\geq 0.
\ee
Eq.(\ref{ae}) must be valid simply because this is the only possible Lorentz covariant expression for the velocity in terms of the local electromagnetic field; and it is clear that, at least where they radiate, the charges move at near the speed of light, and, due to strong radiation overdamping, the charge velocity depends only on the local values of the electric and magnetic fields (see \cite{Gruzinov2013} for further details). 

It remains to calculate the densities $\rho_\pm$. This is almost straightforward, as we know how the charges move:
\be\label{con}
\dot{\rho_{\pm}}+\nabla \cdot (\rho_{\pm}{\bf v}_{\pm})=Q.
\ee
The only subtlety is the pair production rate $Q$. 

For weak pulsars, one postulates pair production near the star with an almost arbitrary prescription which keeps $Q$ positive and large (again compared to Goldreich-Julian per rotation) so long as the proper electric field $E_0$ does not drop well below its typical vacuum value. 

For non-weak pulsars, one needs to add pair production in the radiation zone. Here pairs are produced in photon collisions. The necessary calculation is also clear (see Conclusions of  arXiv:1310.1894) and appears doable.


\begin{thebibliography}{99}

\bibitem{Goldreich1969}
P. Goldreich, W.H. Julian, Astrophys.\ J.\ {\bf 157}, 869 (1969)

\bibitem{Fermi2013}
The Fermi-LAT collaboration, arXiv:1305.4385 (2013)

\bibitem{Gruzinov2013}
A. Gruzinov, arXiv:1309.6974,  arXiv:1310.1894 (2013)






\end{thebibliography}
\end{document}